\documentstyle[epsf,epsfig]{aipproc}
\begin{document}
\newcommand {\be} {\begin{equation}   } 
\newcommand {\ee} {\end{equation}   } 
\newcommand {\bea} {\begin{eqnarray}   } 
\newcommand {\eea} {\end{eqnarray}   } 
\newcommand {\bay } {\begin{array}   } 
\newcommand {\eay } {\end{array}   }
\newcommand {\ol  } {\overline       } 
\newcommand {\lb }  {\label }
\newcommand {\slL}  {\sum_{\ell=0}^L                    }
\newcommand {\nl }  { \newline                           }
\newcommand {\cF  } {{\cal   F }}       
\newcommand {\cH  } {{\cal   H }}       
\newcommand {\cK  } {{\cal   K }}       
\newcommand {\cN  } {{\cal   N }}      
\newcommand {\cZ  } {{\cal   Z }}      
\newcommand {\hsc } {\hspace*{1cm}   }
\newcommand {\fa  } {\forall                         }
\newcommand {\lh  } {\left(                          }
\newcommand {\rh  } {\right)                         }
\newcommand {\lv  } {\left[                          }
\newcommand {\rv  } {\right]                         }
\newcommand {\lgl } {\langle                    }
\newcommand {\rgl } {\rangle                   }
\newcommand {\lc  } {\left\{                         }
\newcommand {\rp  } {\right .                        }
\newcommand {\si  } {\sigma     }
\newcommand {\de  } {\delta     } 
\newcommand {\ep  } {\epsilon   } 
\newcommand {\ta  } {\tau       }
\newcommand {\vsi } {{\vec{\sigma  }}}
\newcommand {\vx  } {{\vec{  x }}}
\newcommand {\Tr  } {\mathop{\mbox{\rm Tr}}   }
\newcommand {\tim } {{\tilde{ m }}}
\newcommand {\tin } {{\tilde{ n }}}
\newcommand {\ev  } {\equiv}
\newcommand {\e   } {\!+\!                  }
\newcommand {\m   } {\!-\!                  }
\newcommand {\noi } {\noindent       }
\newcommand {\ov  } {\over }

\title{\bf Configurational entropy and the one-step RSB scenario
in glasses}

\author{A. Crisanti$^*$, and F. Ritort$^\dagger$} \address{ $^*$
Dipartimento di Fisica, Universit\`a di Roma ``La Sapienza'', and \\
Istituto Nazionale Fisica della Materia, Unit\`a di Roma I \\ P.le Aldo
Moro 2, I-00185 Roma, Italy.\\ $^\dagger$ Physics Department, Faculty of
Physics \\ University of Barcelona, Diagonal 647, 08028 Barcelona, Spain
}

\maketitle

\begin{abstract}
In this talk we discuss the possibility of constructing a fluctuation
theory for structural glasses in the non-equilibrium aging state. After
reviewing well known results in a toy model we discuss some of the key
assumptions which support the validity of this theory, in particular
the role of the configurational entropy and its relation to the
effective temperature. Recent numerical results for mean-field
finite-size glasses agree with this scenario.
\end{abstract}

\section{The glass state}
A theoretical understanding of the physical mechanisms behind the
glass transition remains an open problem \cite{REVIEW}. What is
commonly understood as a glass (for instance, window glass) is a
metastable phase, with free energy higher than that of the
crystal obtained by the continuation of the liquid line below the
melting transition temperature $T_M$. Such a glassy state may be
experimentally achieved by cooling the liquid fast enough. Glasses
share physical properties common to liquids and solids making
difficult to decide in which phase (if any) their are. On the
one hand, atom positions are randomly located in the glass much
alike a liquid and apparently there is no long-range
order. On the other hand, under compression glasses behave like a
solid showing a very low mobility. Glasses constitute a state of
matter in between liquids and solids, not completely classificable as
any of them, which show the following generic features:

\begin{itemize}
\item{\it The viscosity anomaly.} Glasses show a very rapid increase
of the viscosity $\eta$ in a relatively narrow range of
temperatures. For instance, in window glass the viscosity increases
over nearly 20 orders of magnitude by changing the temperature in a
range within a 10\% of the value of the melting transition
temperature. This increase is often well fitted by the
Vogel-Tamman-Fulcher law, $\eta\sim \exp(\Delta/(T-T_0))$ where $T_0$
and $\Delta$ are free parameters. Fragile glasses are those where
$T_0$ is finite while for strong glasses $T_0$ is small (compared to
$T_M$) and the viscosity displays
Arrhenius behavior. By convention the glass transition temperature
$T_g$ is defined such that $\eta(T_g)=10^{13}$ Poise which for many
liquids corresponds to a relaxation time of several minutes .

\item{\it Two-step relaxation.} In the undercooled regime glasses show
relaxation functions with a characteristic two-step form. Intermediate
scattering functions show a first decay to a {\it plateau}
($\beta$-process) followed by a secondary and slower relaxation
($\alpha$-process) which defines the longest activated
time. Although both process are activated their characteristic times
turn out to be well separated below the mode-coupling transition
temperature.

\item{\it Aging.} Suppose a glass is quenched to a temperature $T_f$ below the
glass transition temperature $T_g$. If one were able to measure the
viscosity as a function of time one would observe that it
grows with time $t$ according to the approximate law: $\eta\sim \eta_0
+at$ where $\eta_0$ is the initial value reached soon after the quenching and
$a$ is a temperature dependent parameter $a\sim \exp(-B/T_f)$ where $B$ is
an activation barrier. This growth can be extremely slow since 
$a$ can be very small depending on the
value of the ratio $B/T_f$. The growth of the viscosity 
manifests itself as a dependence of intermediate scattering functions on
the time of the measurement (usually called {\it waiting time} $t_w$)
where the longest decorrelation time depends on the value of the
viscosity $\eta$ at $t_w$. The dependence of the response of the
system in the presence of an external perturbation, 
which softens as time goes by, is commonly referred as {\it
aging}. 

\end{itemize}

\section{A simple solvable model of a glass}
Before describing the key assumptions to understand glasses it is
convenient to consider an instructive example. One of the simplest
solvable models which shows key features of structural glasses is the
oscillator model with parallel dynamics introduced by one of us in
\cite{BPR}. This model contains some essential features of
non-equilibrium thermodynamics applied to glassy systems. The model
is defined by a set of non-interacting $N$ harmonic oscillators with energy:
\be
E=\frac{K}{2}\,\sum_{i=1}^N\,x_i^2,~~~~-\infty < x_i< \infty
\label{eq1}
\ee
where $K$ is a coupling constant. An
effective interaction between oscillators is introduced through a
parallel Monte Carlo dynamics characterized by small jumps $x_i\to
x'_i=x_i+\frac{r_i}{\sqrt{N}};1\le i \le N$ where the variables $r_i$ are
extracted from a Gaussian distribution
$P(r)=\frac{1}{\sqrt{2\pi\Delta^2}}\exp(-r^2/(2\Delta^2))$. At each Monte Carlo
step all oscillators are updated following the previous rule and the
move is accepted according to the Metropolis algorithm with probability
$W(\Delta E)=\min[1,\exp(-\beta \Delta E)]$ where $\Delta E=E(\lbrace
x'\rbrace)-E(\lbrace x \rbrace)$. The smallness of the jumps in the
variables $x_i$ is required in order to give a finite change in the
energy so the acceptance does not vanish in the $N\to\infty$ limit. 

This model (as well as some modifications proposed afterwards
\cite{THEO1}) has been extensively studied and shows the following
scenario. Because of the finiteness of the moves the ground state
cannot be reached in a finite amount of time. Despite the absence 
interactions in the Hamiltonian, the Monte Carlo dynamics induces
entropy barriers corresponding to the flat directions in
energy space that the system hardly finds when the acceptance is
low. A simple calculation at finite temperature shows activated
behavior for the relaxation time $\tau_{relax}\sim
\exp(\frac{K\Delta^2}{8T})$ despite the absence of 
energy barriers in the model (the potential is a single well in $N$
dimensions) making the dynamics of this model quite reminiscent (but
simpler) of that of the Backgammon model \cite{RIT}.

The interesting dynamics in this model is found when studying the
relaxation after quenching to zero temperature. In what follows we
summarize the main findings for this case:
\cite{BPR}:

\begin{itemize}

\item{\it Slow decay of the energy:} The evolution equation for the
energy is Markovian. This simplicity allows for an asymptotic large-time
expansion. The energy asymptotically decays logarithmically $E(t)\sim
1/\log(t)$ and the acceptance ratio decays faster $A(t)\sim 1/(t\,\log(t))$.




\item{\it The effective temperature.} The fluctuation-dissipation (FDT)
ratio \cite{X} can be exactly computed and depends only on the lowest
time. At zero temperature one gets in the large $s$ limit,
\be
T_{\rm eff}(s)=\frac{{\partial C(t,s)}/{\partial s}}{G(t,s)}\to \frac{2E(s)}{N}
\label{eq3}
\ee
i.e., equipartitioning is obeyed off-equilibrium allowing to define an
effective temperature in terms of the FDT ratio. 

\item{\it The role of the configurational entropy.} The effective
temperature previously obtained allows for a definition of a time
dependent configurational entropy.  At time $t$ the number
of configurations explored by the system is given by the surface of an
$N$-dimensional sphere of radius $R=E^{\frac{1}{2}}$ where $E$ is the
energy at time $t$. The configurational entropy is then given by
$\Omega\sim R^{N-1}=E^{\frac{N-1}{2}}$ leading to an extensive
configurational entropy $S_c=\log(\Omega)=\frac{N}{2}\log(E)$
which satisfies, using (\ref{eq3}), the canonical thermodynamic
relations,

\be
\frac{\partial S_c}{\partial E}=\frac{N}{2E}=\frac{1}{T_{\rm eff}}
\label{eq4}
\ee

\item{\it Heat-flow driven by the effective temperature.} Following
\cite{CKP} we can obtain the resulting heat-current by coupling two
identical harmonic oscillator systems with a term $\epsilon x_i y_i$
and measuring the energy flow to order $\epsilon^2$. One gets
\cite{GR},

\be
J(t)=\lim_{t\to s}(\frac{\partial Q(t,s)}{\partial t}-\frac{\partial Q(t,s)}{\partial s})
\label{eq5}
\ee with $Q(t,s)=\frac{1}{N}\sum_{i=1}^N\,x_i(t)y_i(s)$. If one of the
two systems acts as a thermometer with characteristic frequency
$\omega\gg 1/t$ then the measured temperature coincides with the
effective temperature and the Onsager relation is satisfied,

\be
J=L_{QQ}\nabla \bigl (\frac{1}{T}\bigr )
\label{eq6}\ee

with $L_{QQ}\simeq {1}/{(t\,\log(t))}$ and $\lambda\simeq
{\log(t)}/{t}$ for the thermal conductivity \cite{GR}. This example
generalizes the Fourier law (and by extension, all linear relations
between fluxes and forces initially derived in the vicinity equilibrium
\cite{MAZUR}) to the case where fluctuations are present around the
off-equilibrium aging state.

\end{itemize}

\section{The canonical thermodynamic scenario for glasses}

In the previous example we have seen how equipartitioning in the
off-equilibrium state is satisfied when the acceptance rate is
small. This can be easily understood: at zero temperature only
configurations with the same or lower energy are accepted. Because the
size of the moves $\Delta$ is finite the system samples all the
configurations of equal or lower energy with the same
probability. Being their number a monotonically
increasing function of the energy and because  the energy cannot increase at
$T=0$ we conclude that the relevant fluctuations are those which explore
the constant energy surface. After quenching to $T=0$ and for long times, the
probability to explore a different state with energy $E<E^*=E(t)$ is
uniform for all states with a given energy and given by,

\bea
P\sim \Omega(E)=\exp(N\,S_c(E))\sim
\exp\left(\frac{N}{2}\log(E)\right)=\nonumber\\
\exp
\left(\frac{N}{2}\log(E^*)+\frac{N}{2E^*}(E-E^*)\right)
\propto \exp\bigl (\beta_{\rm eff}(E-E^*)\bigr)\,\Theta(E^*-E)~~~~.
\label{eq7}
\eea 
Fluctuations are then described in terms of an effective temperature
given by the configurational entropy as described in
eq.(\ref{eq4}). In the last years Th. M. Nieuwenhuizen, inspired by
results for this oscillator model as well as for the $p$-spin spherical
spin glass has proposed that a similar scenario could apply for generic
glassy systems \cite{THEO2}. Although it is not clear in what conditions these
results are valid for realistic systems the idea is suggestive enough to
merit a more detailed investigation. A more elaborate picture along the
same line has been presented by Franz and Virasoro who, using well known
results from spin glasses and assuming a separation of timescales, have
proposed that a modified version of the Onsager hypothesis should put
the configurational entropy into the game \cite{FV}. Although their analysis is
based on some mean-field models it can be probably
extended to realistic systems under some general mild assumptions.

The idea goes as follows. Imagine a glass that is obtained by quenching
the liquid to a temperature $T_f$ below $T_g$. As already discussed, one
of the main experimental facts regarding relaxational processes (for
which mode-coupling theory gives a fairly good description in a large
range of temperatures) is the emergence of well separated timescales
(the $\alpha$ and $\beta$ processes). Dynamics can then be viewed as if
the system were jumping among different metastable states or {\it
basins}, each basin defined as the set of configurations explored by the
system during a time $t^*$ of the order of the relaxation time
itself. The definition of the precise value of $t^*$ does not matter as
soon as the timescales separation becomes very strong, a condition which
emphasizes the robustness of the present scenario. Fluctuations around a
given configuration have two contributions: 1) Thermal fluctuations {\em
within} a basin, i.e., typical fluctuations occurring on a timescale less
than $t^*$ and driven by the temperature $T_f$ and 2)
Activated jumps between different basins typically occurring for
times larger than $t^*$. If basins with a given free
energy are more or less similar (they are not too big or too small) then
their typical lifetime is of order $t^*$ implying that the probability
to visit them at subsequent times $t>t^*$ is simply proportional to
their number $\exp(S_c(F))$ where $S_c(F)$ defines the configurational
entropy,

\bea P(F)\propto \Omega(F)=\exp\Bigl(S_c(F)\Bigr)= \exp\Bigl(
S_c(F^*)+(\frac{\partial S_c(F)}{\partial
F})_{F=F^*}(F-F^*)\nonumber\\+O(F-F^*)^2\Bigr)\propto \exp\Bigl(
(\frac{\partial S_c(F)}{\partial F})_{F=F^*}(F-F^*)\Bigr)
\label{eq8}
\eea
where $F^*$ is the {\it threshold} value of the free energy, i.e., the
free energy of the basins at time $t$. Note that the present threshold
we define here is conceptually different from that defined in mean-field
$p$-spin models since now $F^*$ is a time dependent quantity. From
(\ref{eq8}) we can infer that the probability distribution for
basin-to-basin fluctuations is given by the configurational entropy
$S_c(F)$ which satisfies the relation,

\be \frac{1}{T_{\rm eff}}=\Bigl (\frac{\partial S_c(F)}{\partial
F}\Bigr )_{F=F^*}
\label{eq9}
\ee

\noindent
and defines an effective temperature for interbasin fluctuations. The
essentials of the physics contained in eq.(\ref{eq8}) is the same as
that contained in eq.(\ref{eq7}) for the oscillator model with the
difference that in the oscillator model fluctuations above the threshold
$E^*$ are forbidden because $T_f=0$ while now fluctuations are dominated
by jumps to basins with free energies around the threshold $F^*$
and differing by a finite amount. We must stress that eq.(\ref{eq8})
gives a good description of the fluctuations when $T_f$ is much smaller than
 $T_{\rm eff}$, otherwise thermal activation driven by $T_f$ will modify
the simple entropic contribution entering in (\ref{eq8}).
Biroli and Kurchan
\cite{BK} have recently addressed the notion of configurational entropy
in the framework of the Gaveau-Schulmann formalism showing that $S_c$ is
a time-dependent quantity. This is in complete agreement with the
present scenario: basins with free energies $F$ above the time-dependent
{\em threshold} $F^*$ are completely washed out as time goes leading to
the appearance of a time-dependent cutoff
$S_c(F)=S_c(F)~~F<F^*$ and $S_c(F)=-\infty~~~F>F^*$ with $S_c(F)$ a monotonous
increasing function of $F$. This keeps the full meaning of (\ref{eq9}).

The conclusion of this discussion is that, as soon as the formula
(\ref{eq8}) gives a faithful description of
basin-to-basin fluctuations occurring beyond a timescale $t^*$ (which
changes itself as a function of time), then it is possible to write down
a dynamical free energy which controls both fast and slow processes:
\be
F_{dyn}=F_{basin}-T_{\rm eff}\,S_c(F)~~~.
\label{eq10}
\ee
Keeping in mind that both terms contribute to physically different
processes occurring in different (well separated) timescales this
corresponds to the formula obtained in the framework of $p$-spin
models. We must stress that in models where free energy basins are not
{\em uniformly} sampled the simple results (\ref{eq9},\ref{eq10}) cannot
hold. Examples of such a class of systems are generic coarsening
models into a finite number of absorbing states \cite{CRRS}.

\begin{figure}
\centerline{\mbox{\epsfxsize=9cm\epsffile{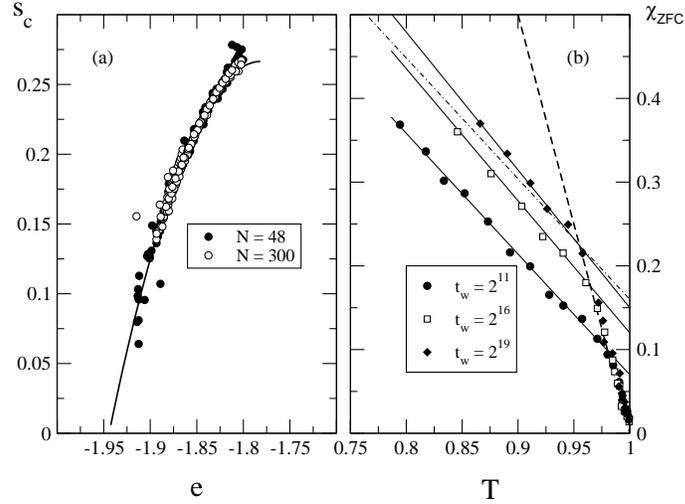}}} 
\caption{(a) Configurational entropy as a function of energy.
         The data are temperatures $T=0.4$, $0.5$, $0.6$, $0.7$,
         $0.8$, $0.9$ and $1.0$.The line is the
         quadratic best-fit. From Ref.\protect\cite{CR1}.
(b) FDT plot for the ROM model (Integrated response function as a
function of IS correlation function).  The dash line has slope
$\beta_{\rm f}= \frac{1}{T_f}= 5.0$, while the full lines is the
prediction $S_c(E)$ from Ref.\protect\cite{CR2}: $T_{\rm eff}(2^{11})\simeq
0.694$, $T_{\rm eff}(2^{16})\simeq 0.634$ and $T_{\rm eff}(2^{19})\simeq
0.608$.  The dot-dashed line is $\beta_{\rm eff}$ for $t_{\rm w}=2^{11}$
drawn for comparison.}
\label{fig}
\end{figure}

This formula has been recently checked in finite-size mean-field
models for glasses where activated processes occur with finite
probability \cite{CR1,CR2}. One can evidenciate the presence of an
activated regime different from that observed in the $N\to\infty$
where ergodicity breaks below the threshold. To verify the above
scenario is necessary to estimate the configurational entropy. A good
recipe to do that is to partition the phase space into basins by
counting the number of energy minima. These are also called inherent
structures and were proposed by Stillinger and Weber as a powerful
tool to investigate the landscape energy surface of liquids \cite{SW}.
The above scenario has been verified in the $ROM$ model \cite{ROM}
which is a faithful microscopic realization of the random energy model
introduced by Derrida \cite{DERRIDA}.  In this model, $F_{basin}\simeq
E_{basin}$ because the intra-basin entropy is very small ($q_{EA}\sim
0.96$ for the metastable states) and the effective temperature can
then be obtained using eq. (\ref{eq9}) with $S_c(F)\simeq S_c(E)$. The
results for the configurational entropy and the effective temperature
(obtained through FDT plots) are shown in figure~(\ref{fig}), panels
$(a)$ and $(b)$ respectively. The agreement between equation
(\ref{eq9}) and the numerics is excellent. Recent results \cite{ST} in
the case of Lennard-Jones glasses, where the contribution $F_{basin}$
turns out to be important, also confirm these results.

The simplicity of the hypothesis presented here suggests that some
generic and simple principles are behind the statistics of
fluctuations around the aging state for structural glasses. Surely we
will see fast developments in the forthcoming years which will 
show to what extent these hypothesis are correct.

{\bf Acknowledgements} We are grateful to A. Garriga for a careful
reading of the manuscript.

\end{document}